\documentclass[a4paper,11pt]{amsart}
\usepackage[utf8]{inputenc}
\usepackage{amsmath,amssymb,amsfonts}
\usepackage{epsfig}
\usepackage{mathrsfs}
\usepackage{graphicx,color}
\usepackage{eucal}
\usepackage{bbm}

\newtheorem{lma}{Lemma}[section]

\newtheorem{alg}{Algorithm}[section]

\newcommand{\norm}[1]{\left| \! \left| #1\right| \!\right|}
\newcommand{\bnorm}[1]{\big| \! \big| #1\big| \!\big|}

\newcommand{\Ex}{\mathbb{E}}
\newcommand{\ip}[1]{\left< #1 \right>}

\newcommand{\Cov}{\textup{Cov}}

\newcommand{\bm}[1]{ \begin{bmatrix} #1 \end{bmatrix}}

\numberwithin{equation}{section}
\font\eka=cmex10
\usepackage{ae}
\def\ind{\mathrel{\hbox{\rlap{%
\hbox to 7.5pt{\hrulefill}}\raise6.6pt\hbox{\eka\char'167}}}}
\begin{document}
%\title[]{Identifying a sparse continuous time dynamics matrix from time series data}

\title[]{A MCMC strategy for identifying a sparse continuous time dynamics matrix from time series data}

\title[]{A MCMC strategy for identifying a sparse continuous time dynamics matrix from time series data}

\title[]{Identifying a sparse continuous time dynamics matrix from time series data by a MCMC approach}

\title[]{Variable selection in linear dynamical systems by Markov Chain Monte Carlo approach}

\title[]{Variable selection in linear dynamical systems by an MCMC approach}

\title[Variable selection in dynamical systems]{Bayesian variable selection in linear dynamical systems}

%Short title inside [ ]

\author[A. Aalto and J. Gon\c{c}alves]{Atte Aalto and Jorge Gon\c{c}alves \vspace{3mm} \\ L\lowercase{uxembourg} C\lowercase{entre} \lowercase{for} S\lowercase{ystems} B\lowercase{iomedicine} \\ U\lowercase{niversity of} L\lowercase{uxembourg}} 
%\address{Luxembourg Centre for Systems Biomedicine, University of Luxembourg; 6~avenue du Swing; 4367 Belvaux; LU} 

\begin{abstract}

%We consider the problem of retrieving the sparsity pattern of the dynamics matrix of a noise-driven linear dynamical system from time series data with low sampling frequency. The low sampling frequency problem is taken into account by treating the continuous-time trajectory as a latent variable, for which a MCMC sampler is constructed. Thus the need to interpolate data is avoided.

%1) the context of your research (whether the scientific or the practical context);

We develop a method for reconstructing regulatory interconnection networks between variables evolving according to a linear dynamical system.
%2) the research problem that motivated the research;
The work is motivated by the problem of gene regulatory network inference, that is, finding causal effects between genes from gene expression time series data.
%3) Research "question"
In biological applications, the typical problem is that the sampling frequency is low, and consequentially the system identification problem is ill-posed.  The low sampling frequency also makes it impossible to estimate derivatives directly from the data.
%4) the aim of the research;
%5) the methodology;
We take a Bayesian approach to the problem, as it offers a natural way to incorporate prior information to deal with the ill-posedness, through the introduction of sparsity promoting prior for the underlying dynamics matrix. It also provides a framework for modelling both the process and measurement noises. We develop Markov Chain Monte Carlo samplers for the discrete-valued zero-structure of the dynamics matrix, and for the continuous-time trajectory of the system.

%6) the results;
%7) the practical implications;
%8) the theoretical implications or implications for further research.

\medskip


%\smallskip
%\noindent
%{\it 2010 AMS subject classification: } 
\end{abstract}

\maketitle

\section{Introduction}

We consider the problem of retrieving the sparsity pattern of the dynamics matrix $A$ in the system
\begin{equation} \label{eq:SDE}
dx=Ax \, dt + du, \qquad x(0)=x_0,
\end{equation}
from time series data $y_j=x(t_j)+v_j$. Here $u$ is an unknown noise process modelled as a Brownian motion with incremental covariance $Q$, and $v_j$'s are measurement noise terms. An additional, deterministic input can be treated by superposition. Our motivation for this problem arises from the field of systems biology, where a topical problem is finding the interconnection network structure between different species. More specifically, we are interested in reconstructing \emph{gene regulatory networks} from gene expression time series data \cite{GRN_review}. In this application, data collection is expensive and laborious, and therefore the temporal resolution tends to be relatively poor and the overall length of the time series short. Consequently, the problem is ill-posed, and additional information needs to be incorporated in order to obtain reasonable solutions. A typical resolution to the identifiability issues is to look for sparse matrices $A$.

Let us discuss first the related problem of variable selection in linear regression, that is, finding the zero-structure of the matrix $A$ from input-output data $\{x_j,y_j\}_{j=1}^N$ connected by
\begin{equation} \label{eq:linreg}
y_j=Ax_j+v_j.
\end{equation}
A sparse solution could be obtained by solving the cardinality-penalised least squares problem, that is, minimising $\gamma|A|_0+\sum_{j=1}^N \norm{y_j-Ax_j}^2$, where $|A|_0$ gives the number of non-zero entries in $A$. However, the cardinality penalty is non-convex and moreover, the problem becomes combinatorial in nature, as each variable combination must be tested separately. A typical remedy is to resort to convex relaxation, that is, penalising instead for the 1-norm of the matrix $A$, defined by $|A|_1=\sum_{i,j=1}^n |A_{i,j}|$. This approach is generally known as Lasso \cite{lasso}. The Lasso approach can also be interpreted as a maximum likelihood estimate assuming Gaussian noise, and Laplace priors for the parameters. This property has been exploited in the so-called Bayesian Lasso approach \cite{BayesLasso},\cite{LassoMCMC}. However, in the Laplace distribution, the value zero is the maximum likelihood estimate, but even there the probability that the coefficient would be zero in a single realization is zero. Therefore the Lasso only produces sparse solutions when it is used in the maximum likelihood estimates.  Sparse realizations can be obtained by introducing a prior distribution for the parameters that have a point mass at zero, leading to a probabilistic counterpart of the cardinality-penalty setup. Clearly the combinatorial nature of the problem remains, but to some extent this can be overcome by using an MCMC strategy. Such strategies tend to spend more time in the ``neighborhood" of solutions with high probability. Variable selection methods based on such probabilistic consideration are considered for example in \cite{spikeslab} introducing so-called ``spike and slab" priors consisting of a mixture of a point mass at zero, and a uniform distribution around zero. Indicator variables  are introduced in \cite{var_selection} and \cite{George}. The article \cite{var_selection} also discusses different types of global priors for the indicator variable. This means that the probability of a certain regression coefficient being zero depends on how many of the other coefficients are zero. Different methods are reviewed and compared in \cite{var_selection_review} and \cite{Dellaportas}.

Comparing the dynamical system \eqref{eq:SDE} and the linear regression case \eqref{eq:linreg}, the main difference is that in \eqref{eq:SDE}, the ``input" on which matrix $A$ operates is the trajectory $x$, and the ``output" is its derivative $\frac{dx}{dt}$. This means that also the input is unknown. In addition, the derivative cannot be estimated directly from the samples $y_j$ due to the low sampling frequency. One approach to tackle this problem is taken in \cite{Kojima} and \cite{Chen_thesis} where the Lasso approach is combined with a Kalman smoother estimating the latent trajectory. The result is an EM type algorithm alternately updating the latent trajectory and the matrix $A$. Another approach is presented in \cite{logA} which is based on a discrete time approach studying $e^{A\Delta T}$ and imposing sparsity on the matrix logarithm. The approach taken here is to impose a sparsity promoting prior probability distribution for the matrix $A$, which --- together with the process noise model --- gives rise to a probability measure for the continuous trajectory $x$. A MCMC sampler is then constructed for both the matrix $A$ (or, more precisely, its zero-structure) and the trajectory $x$. %The MCMC approach also yields more information on the statistics of the zero-structure of $A$, than what is obtained by a maximum likelihood estimate given by the EM-Lasso approach.
The problem of low sampling rate is addressed by sampling the full trajectory $x$, as opposed to sampling only $x(t_j)$. The prior for $A$ is defined through an indicator variable as in \cite{var_selection}, and separate  priors are employed for the indicator variable, and the magnitudes of the non-zero values. A discrete-valued Markov chain is defined that is moving between different indicator variables, that are controlling the zero-structure of $A$. This approach bears some resemblance to the Reversible Jump MCMC which is designed in \cite{RJMCMC} for sampling from distributions with varying dimension. However, assuming a normal distribution for the non-zero elements of $A$, it is possible to integrate out the magnitude parameters, thus avoiding the need to deal with varying dimension of the parameter space.

%The continuous-time system \eqref{eq:SDE} can be discretised to obtain (assume $t_j-t_{j-1}=\Delta t$ for all $j$)
%\[
%x(t_j)=e^{A\Delta t}x(t_{j-1})+w_j
%\]
%where $w_j \sim N\big(0,\int_0^{\Delta t} e^{As}Qe^{A^{\top}s}ds\big)$. If $\Delta t$ is small, it holds that $e^{A\Delta t} \approx I+\Delta tA$, and then it is possible to consider a discrete-time approach and use this approximation to relate the continuous- and discrete-time system matrices. Unfortunately, when the sampling time is long, then sparsity of the continuous-time $A$ does not imply even approximate sparsity of  the discrete-time $e^{A\Delta t}$. The main challenge addressed in this article is the incorporation of the sparsity requirement on the continuous-time level when the sampling frequency is low.

%In this article, the problem is considered as a variable selection problem from a probabilistic perspective, and we introduce a Markov Chain Monte Carlo (MCMC) approach for the task. This approach can be considered as a dynamical variant of the  so-called Bayesian Lasso approach presented in~\cite{LassoMCMC}. They propose an MCMC approach for a variable selection problem in linear regression, that is, finding the zero-structure of matrix $A$ from given input-output data in the form \eqref{eq:linreg} where the inputs $x_j$ are known. In our setup \eqref{eq:SDE}, the ``input" on which matrix $A$ operates is the trajectory $x$, and the ``output" is its derivative $\frac{dx}{dt}$. The trajectory $x$ is considered as a latent variable, for which a sampler is constructed.

The proposed approach gives rise to some challenges related to the MCMC sampling. The continuous-time trajectory of the system is an infinite-di\-men\-sio\-nal random variable. We will employ a Crank-Nicolson sampling scheme for the trajectory in order to achieve high acceptance rates in the sampler. A mixture of discrete and continuous variables is prone to multimodality problems. This problem is addressed by employing a parallel tempering scheme. The outline of the paper is as follows: in order to best convey the main idea, the sampling scheme for the zero-structure of the matrix $A$ is first introduced in the simpler context of variable selection in linear regression. This is the topic of Section~\ref{sec:basic}. The case of dynamical systems is treated in Section~\ref{sec:dynamic}, where we also introduce slight improvements and generalisations of the method involving higher order dynamics. Finally, in Section~\ref{sec:example}, we present a numerical example where the introduced method is compared to the Expectation Maximization (EM) method incorporating a Laplace prior for the elements of matrix $A$, corresponding to the popular Lasso algorithm.

It should be noted that the proposed method is straightforwardly generalizable to nonlinear dynamics $dx = \sum_{j=1}^L w_j \psi_j(x)dt + du$ where $\{ \psi_j \}_{j=1}^L$ is a library of selected nonlinear functions, and their weights $w_j$  are to be determined. A similar approach has been presented in  \cite{SINDy} and \cite{dyn_model_selection}.

%\subsection{References}
%Cite \cite{Klauder} for integrals over infinite dimensional domains, \cite{LassoMCMC} for a similar approach (check also references [6,11,20,25,27] there), and \cite{Beskos_CN,CN_MCMC} for the Crank-Nicolson sampling (Beskos et al. is apparently the one introducing it, the second one is for some general recipies and numerical experiments). {\bf Needed: GRN, RJMCMC, LASSO}
%New: \cite{Dellaportas} \cite{George} \cite{dyn_model_selection} \cite{Chen_thesis} \cite{SINDy}

%\subsection{notation} I'm not sure if there will actually be a notation section. Now it is here just as a quick reference collection and as a reminder to check in the end that all definitions are in the paper.
%\begin{itemize}
%\item $|h|_0$ is the number of non-zero elements in the vector $h$.
%\item $h[S_i]$ etc.
%\item $p(\cdot)$ is used as a generic symbol for a probability distribution when there is no risk of confusion.
%\end{itemize}

\section{Variable selection in linear regression} \label{sec:basic}

In this section we introduce our sampling scheme for sampling the zero-structure of the matrix $A$ in connection of a linear regression problem. Say we have data of input-output pairs $\{x_j,y_j\} \in \mathbb{R}^n \times \mathbb{R}^m$ for $j=1,...,N$ of the form
\[
y_j=Ax_j+v_j,
\]
where $v_j \sim N(0,R)$ and $v_j \perp v_k$ if $j \ne k$. The task is to identify the matrix $A \in \mathbb{R}^{m \times n}$ for which we have prior information that it should be sparse. %Typically such tasks are tackled by formulating the problem as an optimization problem, such as least squares problem $\sum_{j=1}^N\norm{y_j-Ax_j}^2$ with a penalty term for the cardinality of $A$. Unfortunately, with the cardinality penalty the problem is no longer convex, and even further, it is an NP-hard combinatorial problem. Therefore the  problem is typically relaxed by incorporating an $\ell^1$ type penalty for $A$, resulting in the so-called LASSO algorithm.

Let us introduce some notation:
\[
Y=[y_1,...,y_N], \qquad X=[x_1,...,x_N].
\]
\[
[A]_{ij}=h_{ij}s_{ij}, \quad \textup{where } a_{ij} \in \mathbb{R}, s_{ij} \in \{0,1 \},
\]
that is, $s_{ij}$ is a variable indicating whether the $(i,j)$ element of the $A$ matrix is non-zero, and $h_{ij}$ is the magnitude variable. Denote the indicator matrix by $S$, that is $[S]_{ij}=s_{ij}$, and $[H]_{ij}=h_{ij}$. %So with Matlab notation, $A=S \, .*H$. 
For a vector $z \in \mathbb{R}^n$, the notation $z[S_i]$ stands for the vector in $\mathbb{R}^{|S_i|_0}$ that consists of those elements $z_j$ for which $S_{i,j} =1$. For a matrix $P \in \mathbb{R}^{n \times n}$, the notation $P[S_i]$ stands for the ${|S_i|_0 \times |S_i|_0}$ submatrix of $P$ that consists of the elements  $P_{j,k}$ for which $S_{i,j} =1$ and $S_{i,k} =1$. For a matrix $P \in \mathbb{R}^{n \times m}$ (or $\mathbb{R}^{m \times n}$) the notation $P[S_i]$ stands for the $|S_i|_0 \times m$ (or $m \times |S_i|_0$) matrix that consists of those rows (columns) $j$ for which $S_{i,j}=1$.

We wish to sample from the posterior distribution $p(S|X,Y)$ for which it holds that
\begin{align*}
 p(S|X,Y)&=\int p(S,H|X,Y)dH \\ & \propto \int p(Y|S,H,X)p(S,H|X)dH \\ &=\int p(Y|A,X)p(H|S,X)p(S|X)dH
\end{align*}
where the first line is a marginalization integral, the second line is the Bayes' rule, and the third line follows from the probability chain rule.
For given $A$ and $X$, the output $Y$ is Gaussian, that is, $p(y_j|A,X)=N(Ax_j,R)$. The topology is independent of the input data, so $p(S|X)=p(S)$ which is just the prior probability for the topology. At this point let us assume that $R$ is a diagonal matrix, $R=\textup{diag}(r_1,...,r_m)$.

For the matrix $H=[h_1,...,h_m]^{\top}$ we assume that its rows $h_i^{\top}$ are independent, and $h_i \sim N(0,M_i)$. Then the function $p(Y|A,X)p(H|S,X)=p(Y|A,X)p(H)$ is an exponential function where the exponent is a quadratic function of $H$: 
\begin{align*}
& p(S)\int p(Y|A,X) p(H)dH \\ & = \frac{p(S)}{(2\pi)^{(mN+mn)/2}|R|^{N/2}\prod_{i=1}^m |M_i|^{1/2}} \\ & \quad \times \int \exp\left(-\frac{1}{2}\sum_{j=1}^N \norm{y_j-Ax_j}_{R^{-1}}^2-\frac12 \sum_{i=1}^m \norm{h_i}_{M_i^{-1}}^2 \right)dH.
\end{align*}
This marginalization integral can be computed analytically.
Firstly, integrating over the variables $h_{i,j}$ for which $S_{i,j}=0$ corresponds to the usual Gaussian marginalization integral
\[
\int p(H) d h_{(i,j) \in \{(i,j) | S_{i,j}=0 \}} =\frac1{(2\pi)^{|S|_0/2} }\prod_{i=1}^m\frac{\exp\left(-\frac12 \bnorm{h_i[S_i]}_{M_i[S_i]^{-1}}^2 \right)}{ |M_i[S_i]|^{1/2}} . 
\]
%where $h_i[S_i]$ is a vector in $\mathbb{R}^{|S_i|_0}$ containing only those elements of $h_i$ for which $S_{i,j}=1$. Similarly $M_i[S_i]$ is the submatrix of $M_i$ containing only those rows and columns for which $S_{i,j}=1$.

For the remaining part of the exponent it holds that
\begin{align*}
& \frac12\sum_{j=1}^N \norm{y_j-Ax_j}_{R^{-1}}^2+\frac12\sum_{i=1}^m \bnorm{h_i[S_i]}_{M_i[S_i]^{-1}}^2 \\ & =J_{\min}+\frac12\sum_{i=1}^m\ip{h_i[S_i]-h_{i,\min},\left({\frac1{r_i}\mathbb{X}[S_i]+M_i[S_i]^{-1}}\right)\big(h_i[S_i]-h_{i,\min}\big)}
\end{align*}
where $J_{\min}$ is the minimal value of the quadratic exponent and $h_{i,\min}$ is the vector attaining this minimum. %The matrix $\mathbb{X}[S_i]$ is again the submatrix of $\mathbb{X}$ containing those rows and columns for which $S_{i,j}=1$. 
The minimal value $J_{\min}$ is obtained by straightforward differentiation and it is
%\[
%J_{\min}=\frac12 \sum_{i=1}^m Y_i \big(r_iI+\tilde X_i^{\top}\tilde M_i \tilde X_i \big)^{-1}Y_i^{\top}
%\]
\begin{equation} \label{eq:Jmin}
J_{\min}=\frac12 \sum_{i=1}^m \frac1{r_i}  Y_i \left( I- \frac1{r_i}X[S_i]^{\top} \! \left( M[S_i]^{-1}+\frac1{r_i}\mathbb{X}[S_i]\right)^{\! -1}\!\! X[S_i]^{\top} \right)Y_i^{\top} \!\!
\end{equation}
%\begin{equation} \label{eq:Jmin}
%J_{\min}=\frac12 \sum_{i=1}^m \frac1{r_i}  Y_i \left( I- \frac1{r_i}X[S_i]^{\top}\left( M[S_i]^{-1}+\frac1{r_i}\mathbb{X}[S_i]\right)^{-1}X[S_i]^{\top} \right)Y_i^{\top}
%\end{equation}
where $Y_i$ is the $i^{\textup{th}}$ row of $Y$, that is, the $1 \times N$ vector containing the $i^{\textup{th}}$ components of $y_j$ for $j=1,...,N$.% and $X[S_i]$ is the $|S_i|_0 \times N$ matrix that consists of those rows $j$ of $X$ for which $S_{i,j}=1$.

Finally, the full marginalisation integral is
%\begin{align*}
%&p(S)\int p(Y|A,X) p(H)dH \\ & =\frac{p(S)\exp\left(-\frac12 \sum_{i=1}^m \frac1{r_i}  Y_i \left( I- \frac1{r_i}X[S_i]^{\top}\left( M[S_i]^{-1}+\frac1{r_i}\mathbb{X}[S_i]\right)^{-1}X[S_i]^{\top} \right)Y_i^{\top}\right)}{(2\pi)^{mN/2} \prod_{i=1}^m  \big|M_i[S_i]^{-1}+\frac1{r_i}\mathbb{X}[S_i]\big|^{1/2}|M_i[S_i]|^{1/2}r_i^{N/2}}.
%\end{align*}
\begin{align*}
&p(S)\int p(Y|A,X) p(H)dH \\ & =\frac{p(S)\exp\left(-J_{\min} \right)}{(2\pi)^{mN/2} \prod_{i=1}^m  \big|M_i[S_i]^{-1}+\frac1{r_i}\mathbb{X}[S_i]\big|^{1/2}|M_i[S_i]|^{1/2}r_i^{N/2}}
\end{align*}
where $J_{\min}$ is given in \eqref{eq:Jmin}.

%This form is interesting but not really feasible if $N>n$. In that case the exponent can be written as
%\[
%\sum_{i=1}^m Y_i \big(r_iI+\tilde X_i^{\top}\tilde M_i \tilde X_i \big)^{-1}Y_i^{\top}= \sum_{i=1}^m \frac1{r_i}  Y_i \left( I- \frac1{r_i}\tilde X_i^{\top}\left( \tilde M_i^{-1}+\frac1{r_i}\tilde X_i\tilde X_i^{\top}\right)^{-1}\tilde X_i^{\top} \right)Y_i^{\top}
%\]
%and the determinant as
%\[
%|r_iI+\tilde X_i^{\top}\tilde M_i \tilde X_i|=\big|\tilde M_i^{-1}+\frac1{r_i}\tilde X_i\tilde X_i^{\top}\big||\tilde M_i|r_i^N.
%\]

\subsection{The proposal Markov chain and the Metropolis--Hastings algorithm}
%To define the dynamics of the Markov chain, a proposal distribution $g(\hat S|S)$ has to be defined for a given connectivity matrix $S$. We propose a simple componentwise transition, that is, if $s_{ij}=0$, then $\hat s_{ij}=1$ with probability $p_1$ and $\hat s_{ij}=0$ with probability $1-p_1$. Similarly, if $s_{ij}=1$, then $\hat s_{ij}=0$ with probability $p_2$ and $\hat s_{ij}=1$ with probability $1-p_2$.
%
%The probability ratio of the proposed step and the reverse step is needed for the Metropolis--Hastings algorithm and it is given by
%\[
%\frac{g(S|\hat S)}{g(\hat S|S)}=\big(p_1/p_2 \big)^{|S|_0-|\hat S|_0}.
%\]
%

There is some freedom in how to perform a jump from one connectivity matrix to another, that is, designing the proposal distribution $g(\hat S|S)$. We will employ a simple scheme where we randomly pick an element from $S$, and flip it. That is, draw $(\hat i,\hat j)$  from the uniform distribution on  $\{1,...,n\}^2$ and set 
\[
[\hat S]_{i,j} = \left\{ \begin{array}{cl} [S]_{i,j}, & \textup{if } (i,j) \ne (\hat i, \hat j), \vspace{1mm} \\ 1-[S]_{i,j}, & \textup{if } (i,j) = (\hat i, \hat j). \end{array} \right.
\]
This proposal is symmetric so that $g(S|\hat S)=g(\hat S|S)$.

Another possibility is presented in \cite{LassoMCMC}.  Their strategy is to decide whether to add or remove (or neither) a variable from the active regressor set. Say that the probability for an addition is $p_1$ and probability for a removal is $p_2$, so that $p_1+p_2 \le 1$ holds. With probability $1-p_1-p_2$, the active regressor set is not changed, that is, $\hat S=S$. The ratio of the probabilities of a jump and its reverse is a bit complicated, since one has to take into account the extreme cases, when a removal step removes the last remaining regressor, or when an addition step results in a full matrix $S$. In the end, the ratios are for an addition move $S \to \hat S$:
\[
\frac{g(S|\hat S)}{g(\hat S|S)} = \left\{ \begin{array}{cl} \!\! \frac{p_2(n^2-|S|_0)}{p_1(|S|_0+1)}, & \textup{when } |S|_0 \le n^2-2, \vspace{1.5mm} \\ \frac{1}{p_1n^2},  & \textup{when } |S|_0 = n^2-1, \end{array} \right.
\]
and for a removal $S \to \hat S$:
\[
\frac{g(S|\hat S)}{g(\hat S|S)} = \left\{ \begin{array}{cl} \!\! \frac{p_1 |S|_0}{p_2(n^2-|S|_0+1)}, & \textup{when } |S|_0 \ge 2, \vspace{1.5mm} \\ \frac{1}{p_2n^2}, & \textup{when } |S|_0 = 1. \end{array} \right.
\]
Note that an addition move is not possible if $|S|_0 = n^2$ and a removal is not possible if $|S|_0=0$.

For a given connectivity matrix $S$ we define the Metropolis--Hastings number
\[
P(S):=\frac{g(S|\hat S)}{g(\hat S|S)}\frac{p(S)\exp\left(-J_{\min}\right)}{ \prod_{i=1}^m  \big|M_i[S_i]^{-1}+\frac1{r_i}\mathbb{X}[S_i]\big|^{1/2}|M_i[S_i]|^{1/2}}
\]
where $J_{\min}$ is given in \eqref{eq:Jmin}, and $p(S)$ is the user-defined prior probability for this particular zero-structure. It can be defined, for example, using the full number of non-zero elements, $|S|_0$, or the numbers of non-zero elements on each row, $(|S_1|_0,...,|S_m|_0)$, etc.

In the Metropolis--Hastings MCMC algorithm we use the above procedure to sample a new network topology $\hat S$ from the old topology matrix $S$. The acceptance probability of the new topology is then $\min\big\{1,P(\hat S)/P(S)\big\}$. This algorithm is summarised below:

\begin{alg} \mbox{}

\begin{itemize}
\item  Set $\bar P=0 \in \mathbb{R}^{m \times n}$ and $n_c=0$.
\item Pick an initial topology $S^{(0)}$ and compute $P(S^{(0)})$. 
\item  For $l=1,...,N_{\textup{sample}}$
\begin{itemize}
\item Form $\hat S$ from $S^{(l-1)}$ using the procedure described above.
\item Compute $P(\hat S)$.
\item With probability $\min\{1,P(\hat S)/P(S^{(l-1)})\}$, set $S^{(l)}=\hat S$. Otherwise set $S^{(l)}=S^{(l-1)}$.
\item Compute $\bar P=\bar P+S^{(l)}$.
\end{itemize}
\item Compute $P=\bar P/N_{\textup{sample}}$.
\end{itemize}
\end{alg}

As the number of samples grows, the elements of the matrix $P=\bar P/N_{\textup{sample}}$ tend to the matrix $\Ex (S|Y)$, whose elements are the probabilities with which the corresponding elements of $A$ are non-zero. This algorithm is slightly simplified since a burn-in period or any thinning are not explicitly included.

%\fbox{\parbox[t][11em][c]{0.92\textwidth}{\footnotesize{ {\bf Algorithm} 
%\begin{itemize}
%\item  Set $\bar P=0 \in \mathbb{R}^{m \times n}$ and $n_c=0$.
%\item Pick an initial topology $S_0$ and compute $P(S_0)$. 
%\item  For $j=1,...,M$
%\begin{itemize}
%\item Form $\hat S$ from $S_{j-1}$ using the procedure described above.
%\item Compute $P(\hat S)$.
%\item With probability $\min(1,P(\hat S)/P(S_{j-1}))$, set $S_j=\hat S$. Otherwise \\ set $S_j=S_{j-1}$.
%\item Whenever $j \equiv 0$ (mod $m$),  set $\bar P=\bar P+S_j$ and $n_c=n_c+1$.
%\end{itemize}
%\item Compute $P=\bar P/n_c$.
%\end{itemize}}
%}} 

%\section{Errors-in-variables case}
%
%Assume now that we have data $\{\hat x_j,\hat y_j\}$ where
%\[
%\hat x_j=x_j+u_j \qquad \hat y_j = y_j+v_j
%\]
%and
%\[
%y_j=Ax_j.
%\]
%Again we intend to obtain samples from the posterior $p(A|\hat X,\hat Y)$ but this time we sample from the joint posterior $p(X,A|\hat X,\hat Y)$, for which it holds
%\[
%p(X,A|\hat X,\hat Y) \propto p(\hat Y|X,\hat X,A)p(X,A|\hat X)=p(\hat Y|X,A)p(X|\hat X)p(A).
%\]
%Now everything is exactly as in the earlier case, but we need to construct a sampling also for $X$.
%
%

 %%%%%%
 	        %
 	        	%
 %%%%%%
 		%
 		%
 %%%%%%

\section{Linear dynamical systems} \label{sec:dynamic}
%\begin{itemize}
%\item $j$: time index
%\item $i$: output dimension
%\item $k$: input dimension
%\item $l$: MCMC sample number, used as a parenthesised superscript, e.g., $x^{(l)}$
%\end{itemize}

In this section, we will encounter a number of different indices. To improve readability, we shall use index $j$ exclusively to refer to the time discretisation, $i \in \{ 1,...,n\}$ for the output dimension of the state $x$ and measurement $y$, $k \in \{1,...,n\}$ for the input dimension, and $l=1,2,...$ for numbering the samples in the MCMC scheme --- used as a parenthesised superscript, \emph{e.g.}, the $l^{\textup{th}}$ trajectory sample is $x^{(l)}$.

In this section, we formulate the approach for estimating the zero-structure of a sparse matrix $A$ from time series data $Y=[y_0|...|y_N] \in \mathbb{R}^{n\times (N+1)}$, that is, $n$ is the dimension of one measurement, and $N+1$ is the number of samples in the time series. This data is assumed to arise from discrete measurements of a continuous time trajectory,
\[
y_j=x(t_j)+v_j.
\]
The trajectory $x$ is the solution of
\[
dx=Ax \, dt + dw, \qquad x(0)=x_0 \sim N(m_0,P_0)
\]
where $w$ is a Brownian motion with incremental covariance $Q$, which is assumed to be diagonal, with $[Q]_{i,i}=q_i$.  Again the goal is to obtain the posterior probabilities for different structure matrices $S$. The main difference to the previous section is that now we have an additional unknown variable, namely the trajectory $x$. This trajectory will be treated as a latent variable, which will be sampled as well. What makes things slightly tricky is that $x$ is an infinite-dimensional variable.  In particular, $p(S,H,dx|Y)$ is a probability measure on the augmented variable consisting of the discrete-valued graph topology, the continuous-valued parameters $H$, and the infinite-dimensional trajectories $x$:
\begin{align*}
p(S|Y) &= \iint p(S,H,dx|Y)dH  \\
& \propto \iint p(Y|x)p(dx|S,H)p(S)p(H) dH  \\
& = p(S) \int p(Y|x) \left( \int p(dx|S,H)p(H)dH \right).
\end{align*}
For a background on infinite-dimensional integrals, we refer to  \cite{Klauder} and for background on stochastic processes, see \cite{stroock}.

Given $A$, that is, $S$ and $H$, the trajectory $x$ is a Gaussian process. The measure of the process $x$ is continuous with respect to the Wiener measure $\mathcal{W}_Q$ corresponding to the incremental covariance $Q$. By the Cameron--Martin theorem \cite[Theorem~8.2.9]{stroock}, it holds that
\[
p(dx|S,H) = \exp\left( \int_0^T \ip{Ax,dx}_{Q^{-1}}-\frac1{2}\norm{Ax}_{L^2(0,T;Q^{-1})}^2 \right)\mathcal{W}_Q(dx).
\]
The exponent is a quadratic function of $A$. Therefore, we impose a normal prior to the rows of $H$, that is, $h_i \sim N(0,M_i)$, and, as before in the linear regression case, the integral with respect to $H$ can be computed analytically like in the basic case in Section~\ref{sec:basic}. 
Denote by $\mathbb{X}$ the matrix defined elementwise
\[
[\mathbb{X}]_{i,k}=\int_0^T x_i(t)x_k(t)dt.
\]
The integral then yields
\begin{align*}
&\int p(dx|S,H)p(H)dH \\ & \propto  \prod_{i=1}^n \frac{\mathcal{W}_{q_i}(dx_i)}{\big|M_i[S_i]^{-1}+\frac1{q_i}\mathbb{X}[S_i] \big|^{1/2} \big| M_i[S_i] \big|^{1/2}} \\  & \quad \times \exp\left(\sum_{i=1}^n \frac1{2q_i^2} \big[ x[S_i],dx_i \big]^{\top} \left(M_i[S_i]^{-1}+\frac1{q_i}\mathbb{X}[S_i] \right)^{-1} \big[ x[S_i],dx_i \big] \right) \\
 & =\exp(\Phi(x))\prod_{i=1}^n  \frac1{\big|M_i[S_i]^{-1}+\frac1{q_i}\mathbb{X}[S_i] \big|^{1/2} \big| M_i[S_i] \big|^{1/2}} \mathcal{W}_Q(dx) 
\end{align*}
The bracket notation is defined for a vector-valued function $w \in L^2(0,T;\mathbb{R}^m)$ %for some $m$ 
as a vector in $\mathbb{R}^m$ defined elementwise as the Ito integral
\[
[w,dv]_k=\int_0^T w_k dv(t).
\]
The functional $\Phi(x)$ is defined by
\begin{equation} \label{eq:phi}
\Phi(x):=\sum_{i=1}^n \frac1{2q_i^2} \big[ x[S_i],dx_i \big]^{\top} \left(M_i[S_i]^{-1}+\frac1{q_i}\mathbb{X}[S_i]\right)^{-1}\big[ x[S_i],dx_i \big].
\end{equation}

\subsection{Sampling strategy}

Common MCMC strategies' acceptance probabilities decrease to zero as the dimension of the distribution increases. In particular, this becomes a problem when sampling some discretised infinite-dimensional object, and one wishes to refine the discretisation. We introduce a Crank--Nicolson sampling scheme \cite{Beskos_CN,CN_MCMC} to speed up the sampling. Crank--Nicolson sampling is based on implementing the ``Gaussian part" of the posterior distribution already in the sampling scheme, and then it does not affect the acceptance probability. That is, assume we wish to sample from a distribution that has the form $p(x) \propto \Phi(x)N(x;m,P)$. A Crank-Nicolson sampler draws samples from $N(m,P)$ by
\[
\hat x=m+\sqrt{1-\varepsilon^2}(x^{(l)}-m)+\varepsilon w,
\]
where $w \sim N(0,P)$, and $x^{(l)}$ is the current sample. The acceptance probability of the sample is computed using only the non-Gaussian part of the distribution, that is, $a(\hat x,x^{(l)})=\min\{1,\Phi(\hat x)/\Phi(x^{(l)})\}$.

In our case, the Gaussian part is $p(Y|x)\mathcal{W}_Q(dx)$. However, sampling from this distribution leads to poor performance, since --- loosely speaking ---  the measure $p(Y|x)\mathcal{W}_Q(dx)$ is concentrated on very different area in the infinite-dimensional space of trajectories as the full posterior measure. We wish to design a sampling measure that is proportional to $\mathcal{W}_Q(dx)$, and that is concentrated on the same area as the full posterior.

Let us introduce the used sampling scheme in the infinite-dimensional context. The practical implementation in discretised form will be presented later.
We propose a two-phase sampling scheme where we first sample $\hat Y:=[\hat y_0,...,\hat y_N]$ from the measurement distribution, that is, $\hat y_j \sim N(y_j,R)$. Then define $m[\hat Y] \in H^1(0,T;\mathbb{R}^n)$ as the continuous, piecewise linear (on intervals $[t_j,t_{j+1}]$) function, for which it holds $m[\hat Y](t_j)=\hat y_j$. The trajectory sample is then $\hat x=m[\hat Y]+\hat b$ where $\hat b$ is a collection of $n$ independent Brownian bridges that satisfy $\hat b(t_j)=0$. That is, each component $b_i$ is a Gaussian process with covariance function
\[
\Cov(\hat b_i(t),\hat b_i(s))=\begin{cases} q_i\frac{\big(t_{j+1}-\max(t,s)\big)\big(\min(t,s)-t_j\big)}{t_{j+1}-t_j}, \qquad \textup{when } t,s \in [t_j,t_{j+1}] \vspace{1mm} \\ 0 \hspace{52.5mm} \textup{otherwise}.  \end{cases}
\]

In the following lemma, it is shown that the proposed sampling scheme equipped with a suitable acceptance-rejection mechanism, is indeed equivalent to sampling from the conditioned measure $\mathcal{W}_Q(dx|y)$. Only the one-dimensional case $n=1$ is considered for simplicity of notation. The higher dimensional trajectory samples are just collections of one-dimensional trajectories.
\begin{lma} \label{lma:wiener}
Say the current state trajectory sample is $x^{(l)}$. The sampling scheme described above, combined with Metropolis--Hastings acceptance ratio
\[
a(x^{(l)},\hat Y,\hat b)=\min \left\{ \exp\left( \sum_{j=0}^{N-1} \frac{ \big(x^{(l)}(t_{j+1})-x^{(l)}(t_j)\big)^2-\big(\hat y_{j+1}-\hat y_j\big)^2}{2q(t_{j+1}-t_j)} \right),1\right\}
\]
is equivalent with sampling from the conditioned Wiener measure $\mathcal{W}_q(dx|y)$.
\end{lma}
Note that the result does not depend on how $\hat b$ and $\hat Y$ are sampled. Later, we will construct Crank--Nicolson samplers for both $\hat b$ and $\hat Y$.
\begin{proof}
An equivalent way to sample as described above is to sample first $\hat Y$ and then a Wiener process $w$ with incremental covariance $q$. Then denote $\bar w=[w(t_0),...,w(t_N)]$, and set
\[
\hat x=w+m[\hat Y-\bar w]
\]
and so $d\hat x=m[\hat Y-\bar w]'dt+dw$. The process $m[\cdot]$ belongs to the Cameron--Martin space of the Wiener measure, that is, $H^1$, and so we can use the Cameron--Martin theorem to obtain a measure for the process $\hat x$ with respect to the Wiener measure $\mathcal{W}_q(dw)$:
\begin{align*}
\frac{p(d\hat x|\hat y)}{\mathcal{W}_q(dw)}=&\exp\left(-\frac1q \int_0^T m[\hat y-\bar w]' dw-\frac1{2q} \norm{m[\hat y-\bar w]'}_{L^2(0,T)}^2 \right) \\
=&\exp\Bigg(\frac1q \sum_{j=1}^{N-1} \Bigg[ \frac{y_{j+1}-w(t_{j+1})-(y_j-w(t_j))}{t_{j+1}-t_j}(w(t_{j+1})-w(t_j)) \\
& \qquad \qquad -\frac12 \left( \frac{y_{j+1}-w(t_{j+1})-(y_j-w(t_j))}{t_{j+1}-t_j}\right)^2(t_{j+1}-t_j) \Bigg]\Bigg) \\
=&\exp\Bigg( \frac1{2q} \sum_{j=1}^{N-1} \left[- \frac{(y_{j+1}-y_j)^2}{t_{j+1}-t_j}+\frac{(w(t_{j+1})-w(t_j))^2}{t_{j+1}-t_j} \right]\Bigg)
\end{align*}
Now $\exp\left( \frac1{2q} \sum_{j=1}^{N-1}\frac{(w(t_{j+1})-w(t_j))^2}{t_{j+1}-t_j}\right)\mathcal{W}_q(dw)$ is exactly the described measure for the process $b=w-m[\bar w]$.
\end{proof}

With this sampling scheme, the proposal distribution already takes into account the data fit and the trajectory smoothness between data points. Therefore the acceptance probability does not tend to zero as the discretisation is refined. In the full posterior sampling, the acceptance probability $a(\cdot,\cdot,\cdot)$ given in the lemma is combined with the part arising from the term $Ax$ in the dynamics equations.

For a practical implementation of the scheme, we need a finite-dimensional subspace of $L^2(0,T)$ that contains the piecewise linear functions $m[\hat y]$. Piecewise linear hat functions with a finer discretisation are a natural choice for the basis of the finite-dimensional subspace. Assume now that the output data is sampled with constant sampling frequency and all dimensions of the state are measured at the same times $t_j=j\Delta T$, $j=0,...,N$, and denote $N\Delta T=T$. This assumption is made mostly for clarity of presentation. Divide each interval [$(j-1)\Delta T,j\Delta T]$ to $n_{\textup{step}}-1$ pieces, where $n_{\textup{step}}$ is a design parameter, and denote $\delta T=\Delta T/(n_{\textup{step}}-1)$. The hat functions are defined as
\[
\phi_j(t)=\max\left\{0,1-\frac{|t-j\delta T|}{\delta T}\right\} \qquad j=0,...,Nn_{\textup{step}}, \ t \in [0,T].
\]
Define the matrices $K$ and $L$ elementwise
\[
K_{i,k}:=\int_0^T \phi_i(t)\phi_k(t)dt \quad \textup{and} \quad L_{i,k}:=\int_0^T \phi_i'(t)\phi_k(t)dt.
\]
With the chosen functions $\phi_j$, these matrices are
\[
K=\frac{\delta T}{6} \bm{2 & 1 & & &  \\ 1 & 4 & 1 && \\& \ddots & \ddots & \ddots & \\ &&1&4&1 \\ &&& 1&2 } \quad \textup{and} \quad L=\frac12\bm{-1 & -1 & & &  \\ 1 & 0 & -1 && \\ & \ddots & \ddots & \ddots & \\ &&1&0&-1 \\ &&& 1&1}.
\]
Define also the embedding matrix $P_{\textup{emb}} \in \mathbb{R}^{(N+1)\times (Nn_{\textup{step}}+1)}$ such that for $\hat Y \in \mathbb{R}^{n\times (N+1)}$, the product $\hat YP_{\textup{emb}} \in \mathbb{R}^{n\times (Nn_{\textup{step}}+1)}$ gives $m[\hat Y]$ in the basis $\{\phi_j\}_{j=0}^{Nn_{\textup{step}}}$. For example, with $n_{\textup{step}}=3$, this matrix is
\[
P_{\textup{emb}}=\frac13 \bm{3&2&1&&&&&& \\ &1&2&3&2&1&&&\\ &&&&1&2&3&2&1\\ &&&&&&&&\ddots}.
\]
To sample the Brownian bridge term $\hat b$, we use the Karhunen--Lo\`{e}ve expansion using sinusoidal basis functions. To this end, define the matrix $P_b \in \mathbb{R}^{(n_{\textup{step}}-1)\times (n_{\textup{step}}-1)}$ whose columns consist of the discretised basis functions:
\[
[P_b]_j=\frac{\sqrt{2T_s}}{\pi j}\bm{\sin\left( \frac{1\pi j}{n_{\textup{step}}}\right) \ \sin\left( \frac{2\pi j}{n_{\textup{step}}}\right) \ \dots \ \sin\left( \frac{(n_{\textup{step}}-1))\pi j}{n_{\textup{step}}}\right)}^{\top}.
\]

The resulting sampling scheme is presented in the form of an algorithm.
\begin{alg} \label{alg:basic}
Initialisation:
\begin{itemize}
\item Choose the discretisation level $n_{\textup{step}}$ and the proposal step length parameter $\varepsilon \in (0,1)$.
\item Form $K$, $L$, $P_{\textup{emb}}$, and $P_b$.
\item Choose initial trajectory $X^{(0)}$, and initial topology $S^{(0)}$.
\end{itemize}

Sampling (for $l=1,...,N_{\textup{sample}}$):\begin{itemize}
\item  Sample $\hat S$ from $S^{(l-1)}$ as described in Section 2.
\item Sample $\hat Y=Y+\sqrt{1-\varepsilon^2}(\hat Y^{(l-1)}-Y)+\varepsilon \sqrt{R}G^{(l)}$ where $G^{(l)}$ is an $n \times (N+1)$ matrix whose each element is an independent, normally distributed random variable with zero mean and variance one.
\item Sample $\hat X=\big( \hat Y-\sqrt{1-\varepsilon^2}\hat Y^{(l-1)}\big)P_{\textup{emb}}+\sqrt{1-\varepsilon^2}\hat X^{(l-1)}+\varepsilon B^{(l)}$ where $B^{(l)}$ consists of the Brownian bridges between measurements.
\item Compute $\mathbb{X}=\hat XK\hat X^{\top}$, and $\mathbb{D}=\hat XL\hat X^{\top}-\frac{TQ}{2}I$. The term $\frac{TQ}{2}I$ arises from the Ito integral formula. Denote the $i^{\textup{th}}$ row of $\mathbb{D}$ by $\mathbb{D}_i$.

\item Compute the Metropolis--Hastings number for the new candidate sample
\begin{equation} \label{eq:MH_P}
P(\hat S,\hat X)=p(\hat S)\exp(\Phi(\hat X,\hat S))\prod_{i=1}^n  \frac{\exp\left( -\frac{1}{q_i\Delta T} \sum_{j=1}^N (\hat Y_{i,j}-\hat Y_{i,j-1})^2  \right)}{\big|M_i[\hat S_i]^{-1}+\frac1{q_i}\mathbb{X}[\hat S_i] \big|^{1/2} \big| M_i[\hat S_i] \big|^{1/2}}
\end{equation}
where
\begin{equation} \label{eq:phi_disc}
\Phi(X,S)=\sum_{i=1}^n \frac1{2q_i^2}\mathbb{D}_i[S_i] \left( M_i[S_i]^{-1}+\frac1{q_i}\mathbb{X}[S_i] \right)^{-1} \mathbb{D}_i[S_i]^{\top}.
\end{equation}
Note that $\hat Y$ is not explicitly a variable of $P(\cdot,\cdot)$ because it can be obtained from $\hat X$ by $\hat Y_{i,j}=\hat X_{i,(j-1)n_{\textup{step}}+1}$.
The acceptance probability of the new sample is
\[
\min\big\{1,P(\hat S,\hat X)/P(S^{(l-1)},\hat X^{(l-1)})\big\},
\]
that is, with this probability, set $S^{(l)}=\hat S$, $\hat X^{(l)}=\hat X$, and $\hat Y^{(l)}=\hat Y$. Otherwise, set $S^{(l)}=S^{(l-1)}$, $\hat X^{(l)}=\hat X^{(l-1)}$, and $\hat Y^{(l)}=\hat Y^{(l-1)}$.
\end{itemize}
\end{alg}
Forming the Brownian bridge term $B^{(l)}$ is difficult to present using standard notation, but it is efficiently done using the MATLAB code line
\begin{center}
{\tt 
B=[zeros(n,1),C*reshape([Pb*randn(n1,n2);zeros(1,n2)],[],n)']}
\end{center}
where {\tt n1}$\,=n_{step}-1$ and {\tt n2}$\,=nN$ and {\tt C}$\,=Q^{1/2}$.

Note that in principle there is no reason why the same step size $\varepsilon$ should be used for both $\hat Y$ and $\hat X$. Also, if some other method is used for sampling the indicator matrix $S$, then the proposal ratio $\frac{g(S|\hat S)}{g(\hat S|S)}$ must be included in the Metropolis--Hastings number $P(S,X)$, see Section~2.

\subsection{Alternative Gibbs sampler}

Under the fairly natural assumption that the topology prior $p(S)$ can be factorised with respect to the rows of $S$, that is, $p(S)=\prod_{i=1}^n p_i(S_i)$, then the algorithm can be made more efficient by introducing a Gibbs sampler that is updating first each row of $S$ separately, and then the trajectory $(X,Y)$. Note that the posterior decomposes also with respect to the rows of $S$, and subsequently the Metropolis--Hastings number in \eqref{eq:MH_P} can be factorised to
\[
P(S,X)=\prod_{i=1}^n P_i(S_i,X)
\]
where $P_i(S_i,X)=p_i(S_i)\exp(\Phi_i(X,S))  \frac{\exp\left( -\frac{1}{q_i\Delta T} \sum_{j=1}^N (\hat Y_{i,j}-\hat Y_{i,j-1})^2  \right)}{\big|M_i[S_i]^{-1}+\frac1{q_i}\mathbb{X}[S_i] \big|^{1/2} \big| M_i[S_i] \big|^{1/2}}$, and $\Phi_i(X,S)$ contains simply the $i^{\textup{th}}$ term of the sum in $\Phi(X,S)$ given in \eqref{eq:phi_disc}.

The key steps of Algorithm~\ref{alg:basic} are modified as follows:
\begin{itemize}
\item For $i=1,...,n$
\begin{itemize}
\item Sample the new row $\hat S_i$ from the current sample $S_i^{(l)}$.
\item Accept with probability $\min \{ 1, P_i(\hat S_i,X^{(l)})/P_i(S_i^{(l)},X^{(l)}) \}$.
\end{itemize}
\item Sample $\hat Y$ and $\hat X$ as in Algorithm~\ref{alg:basic}.
\item Accept with probability $\min \{ 1, \prod_{i=1}^n P_i(S_i^{(l+1)},\hat X)/P_i(S_i^{(l+1)},X^{(l)}) \}$
\end{itemize}

Note that in this modification, each factor $P_i(S,X)$ is stored separately.

\subsection{Hyperparameter sampling}

Typically even the hyperparameters $M_i$, $q_i$ and $r_i$ are not known, and they can be sampled as well. Again, sampling the process noise covariance $q_i$ poses an additional technical problem, because the Wiener measures corresponding to different covariances are not equivalent. This means that if nothing else is done, then in the infinitesimal discretisation limit, a step where $q_i$ increases is always accepted, whereas a step where $q_i$ decreases is never accepted. To prevent this, The Brownian bridge term in the trajectory has to be scaled by $\textup{diag}\big(\big(\hat q_i/q_i^{(l)}\big)^{1/2}\big)$. That is, if the current trajectory $X^{(l)}$ is decomposed into $X^{(l)}=Y^{(l)}P_{\textup{emb}}+\big(X^{(l)}-Y^{(l)}P_{\textup{emb}} \big)$, then when hyperparameter $\hat q$ is sampled, then also the trajectory is scaled to obtain a candidate $\hat X=Y^{(l)}P_{\textup{emb}}+\textup{diag}\big(\big(\hat q_i/q_i^{(l)}\big)^{1/2}\big)\big(X^{(l)}-Y^{(l)}P_{\textup{emb}} \big)$, which is accepted if $\hat q$ is accepted. 

We assume that the matrices $M_i$ are assumed diagonal, and they are assumed to be of the form $M_i=m_i M_0$ where $M_0$ is diagonal with 
\[
[M_0]_{i,i}=\left(\frac{(t_1-t_0)^2}{4}Y_{i,0}^2+\frac{(t_N-t_{N-1})^2}{4}Y_{i,N}^2+\sum_{j=1}^{N-1} \frac{(t_{j+1}-t_{j-1})^2}{4}Y_{i,j}^2\right)^{-1}.
\]
The purpose of this choice is to scale all potential regulators to same magnitude so that the scales would not matter in the variable selection.

When the hyperparameters $q$ and $m_i$ are sampled, their acceptance is based on computing the Metropolis--Hastings numbers using these new variables. That is, using for example random walk sampling, $\hat m= m^{(l)}+v$, and $\hat q = q^{(l)}+w$, then these samples are accepted with probability
\[
\frac{P_{\hat m,\hat q}(S^{(l)},X^{(l)})}{P_{m^{(l)},q^{(l)}}(S^{(l)},X^{(l)})}\frac{p(\hat m)p(\hat q)}{p(m^{(l)})p(q^{(l)})}\prod_{i=1}^n \frac{(q_i^{(l)})^{N/2}}{\hat q_i^{N/2}}
\]
where $p(m)$ and $p(q)$ are the user defined hyperpriors. The product-term arises from the Wiener measure factorization in Lemma~\ref{lma:wiener}. It is not necessary to sample $q$ and $m$ simultaneously, and $m$ can even be sampled one component at a time without increasing computational complexity.

The measurement noise variance $r$ can be sampled using the random walk sampling $\hat r = r^{(l)}+u$ and the acceptance probability is given by
\[
\frac{p(\hat r)}{p(r^{(l)})}\prod_{i=1}^n \frac{(r_i^{(l)})^{(N+1)/2}}{\hat r_i^{(N+1)/2}}\exp\left( \frac{\bnorm{Y_i-Y_i^{(l)}}^2}{2}\left( \frac1{r_i^{(l)}}-\frac1{\hat r_i}\right) \!\!   \right)
\]
where again $p(r)$ is a user defined hyperprior. Obviously $r$ can also be sampled one component at a time, since its posterior readily factorizes.

\subsection{Output dynamics} \label{sub:out}

The presented algorithm can be considered as a network identification method in the spirit of \emph{dynamical structure functions} \cite{DSF} or \emph{(module) dynamical networks} \cite{module} with a simplified transfer function structure. Each node in the network consists of a state variable $x_i$ whose dynamics in frequency domain are given by
\[
X_i(s)=G_{i,\textup{in}}(s)\left(\sum_{j=1 \atop j \ne i}^n a_{i,j}X_j(s)+ U_i(s) \right)
\]
where $X_i :=\mathcal{L}(x_i)$, $G_{i,\textup{in}}(s)=\frac1{s-a_{i,i}}$, and $U_i(s)=\mathcal{L}\left(u_i \right)$.
Our motivation for this work arises from gene regulatory network identification problem, where the state variables $x_j$ are gene expression levels. Sometimes the regulatory effect from one gene to another happens through a protein interaction. The concentrations of proteins are (usually) not measured, but we can try to take these potential interactions into account by augmenting the output variables with simple dynamics. The variable that is fed to other dynamics is then
\[
\bm{ G_{i,\textup{out}}(s) \\ I} X_i(s)
\]
where the first component models the hidden protein concentration. The transfer function from variable $j$ to variable $i$ takes either the form $a_{i,j}G_{i,\textup{in}}(s)$ or $a_{i,j}G_{i,\textup{in}}(s)G_{j,\textup{out}}(s)$.

Assuming that also the output dynamics are of first order, the dynamics in state space formalism are governed by
\[
\bm{dx \\ dz}=\bm{A_x & A_z \\ I & D}\bm{x \\ z} + \bm{du \\ 0},
\]
where $D=\textup{diag}\big(\{d_j\}_{j=1}^n\big)$. Earlier we did not impose any stability requirements in the prior for $A$. However, for the unobserved part of the state, we impose a stability condition $d_j \le 0$ for all $j=1,...,n$, obtained through a prior $d_j \sim\frac{1}{l_j}\mathbbm{1}_{\mathbb{R}^-}(d_j)\exp\left(d_j/l_j\right)$. 

From the point of view of the algorithm, it doesn't seem feasible to integrate out the parameters $d_j$. In this case the algorithm proceeds as follows:
\begin{itemize}
\item Sample the trajectory $x$;
\item Sample $d_j$'s;
\item Sample z(0) and compute $z$;
\item Compute $\mathbb{X}$ and $\mathbb{D}$ as before;
\item Continue as in the basic case in Algorithm~\ref{alg:basic}.
\end{itemize}

Notice that the topology matrix is now doubled in size, that is $S \in \mathbb{R}^{n \times 2n}$. It is of course possible to give different priors to different parts of $S$, and it is even possible to impose an exclusive prior such that only one of the output components of one node --- that is, $x_j$ or $z_j$ --- can be used as an input for another state variable. Such prior can be easily encoded using an $n \times n$ topology matrix where each entry has three possible values.

\subsection{Tempered schemes} \label{sec:temp}

A combination of discrete parameter (the topology $S$) and a continuous parameter (the trajectory $x$) can be difficult to sample because the target may be multimodal. One way to accommodate multimodal targets is to use some tempering scheme, such as parallel tempering \cite{Temp}, tempered transitions \cite{multimodal}, or tempered secondary chains \cite{MCMC_2nd}. In the presented sampling strategy, a tempered scheme can be implemented in a straightforward manner. 
%Let us first discuss how tempering is implemented, and then present briefly the parallel tempering scheme, as that seemed to yield the biggest improvement in our simulations.
The only change is that the Metropolis--Hastings number $P(S,X)$ is replaced by $P(S,X)^{\beta}$ when computing the acceptance probability, where $\beta \in (0,1]$ is the \emph{inverse temperature}. %Since part of the posterior distribution is already built in the Crank--Nicolson sampler, it is also possible to implement the higher temperature in the trajectory sampler. It seemed to improve the performance to multiply the first two or three columns of $P_b$ with $\beta^{-1/2}$. Note that if the whole matrix is scaled, then at the infinitesimal discretisation limit, the measures corresponding to different temperatures are not absolutely continuous with respect to each other. This leads to poor performance of any tempering scheme (and to complete failure as the discretisation is refined).

In the parallel tempering scheme, a series of (inverse) temperatures is chosen such that $1=\beta_0>\beta_1>...>\beta_h>0$. Parallel chains --- each with different temperature --- are then run simultaneously. Every now and then, a swap of two states from different chains (with adjacent temperatures) is attempted. The attempted swap of states $(X[j],S[j])$ and $([X[j+1],S[j+1])$ corresponding to temperatures $\beta_{j}$ and $\beta_{j+1}$, respectively, is accepted with probability
\[
\min\left\{1, \frac{P(S[j+1],X[j+1])^{\beta_j}P(S[j],X[j])^{\beta_{j+1}}}{P(S[j],X[j])^{\beta_j}P(S[j+1],X[j+1])^{\beta_{j+1}}}\right\}.
\]
The samples corresponding to the lowest temperature $\beta_0=1$ are collected, and the samples corresponding to higher temperatures are discarded. Thus, better mixing of the Markov chain is gained at the expense of higher computational effort. Deciding the number of parallel chains and the used temperatures typically needs some trial runs and parameter tuning based on the observed acceptance probabilities of the swaps.

%parallel tempering, described in \cite{Temp,multimodal} to account for multimodal posterior distributions. In the parallel tempering scheme the higher temperature should be implemented only in the $Y$-term in the acceptance probability $a(\cdot,\cdot,\cdot)$ and not the Wiener measure in the proposal distribution, or otherwise the high-temperature samples are never accepted in the lower temperature chain (at the infinitesimal discretisation limit). 

The downside of the parallel tempering scheme (and other tempering schemes) is the rather high computational burden, compared to the basic scheme. A fast, heuristic method can be obtained by applying a higher temperature only when sampling a new structure matrix $S$. That is, on the $l^{\textup{th}}$ iteration, the sample $\hat S$ is accepted with probability 
$
\min\left\{ 1, \frac{P(\hat S,X^{(l-1)})^{\beta}}{P(S^{(l-1)},X^{(l-1)})^{\beta}} \right\}
$
and then the sample $\hat X$ is accepted with probability 
$
\min\left\{ 1, \frac{P(S^{(l)},\hat X)}{P(S^{(l)},X^{(l-1)})} \right\}.
$
The results using this heuristic scheme were practically undistinguishable from the results from the parallel tempering scheme in our test problem. Moreover, tuning the parallel tempering scheme is more difficult.

\subsection{Combining several time series}

It is straightforward to combine several time series to obtain one matrix $A$. The changes that need to be made to the algorithm are as follows: 1) Separate samplers need to be constructed for each of the continuous time trajectories; 2) The matrices $\mathbb{X}$ and $\mathbb{D}$ consist of sums of the corresponding matrices for the single trajectories; 3) In the acceptance probability, one must include the $\hat Y$ terms for all time series separately.

If necessary, it is also possible to improve the acceptance probabilities by updating only one trajectory at a time, although in that case one needs to further factorise the target distribution with respect to the different time series. If the time series are similar in terms of sampling frequency and number of samples, the task is somewhat easier since it is possible to use same samplers for all trajectories.

\section{Numerical example} \label{sec:example}
\begin{figure}[t]
\vspace{-1.5cm}
\includegraphics[width=11cm]{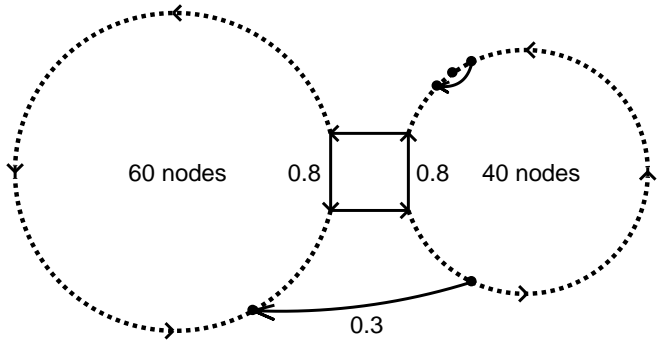}
\vspace{-1.5cm}
\caption{The graph corresponding to the connectivity of matrix $A$. Dotted lines represent high number of nodes connected by edges all pointing at the same general direction.}
\label{fig:graph}
\end{figure}

The numerical example treats a transport-type system with state space $\mathbb{R}^{100}$. The graph corresponding to the ground truth matrix is sketched in Figure~\ref{fig:graph}. The structure consists of two connected rings with 40 and 60 nodes. These nodes are not explicitly shown in the figure. An edge in the graph corresponds to a value one in the matrix $A$, unless another value is given in the graph. The diagonal values are set so that each column sum is zero, corresponding to a transport-type system where the 1-norm of solutions is conserved (in the noise-free case). Altogether there are 10000 entries in the matrix $A$ and 204 of them are non-zero.

Trajectories were simulated from this system, each starting from an initial state that was drawn from a normal distribution $N(0,2^2I)$. The lengths of the trajectories were 10 time units and the sampling frequency was 0.5 or 1. The time series therefore consisted of 21 or 11 samples each. A small process noise was added to each dimension of the state space. These process noises were independent realizations of the Ornstein--Uhlenbeck process
\[
du=-10u \, dt + dw
\]
where $w$ is a Brownian motion with incremental covariance 4. Independent noise terms were added to each measurement, that were drawn from the normal distribution $N(0,0.04^2)$.

Hyperparameter sampling was used in our method. For $q_i$ and $r_i$, the noninformative inverse Gamma prior was used, and for the magnitude variance parameter $m_i$, we used a prior 
\[
p(m_i) \propto \frac{m_i}{\textup{Var}(Y_i)}\left(20-\frac{m_i}{\textup{Var}(Y_i)}\right)\exp\left(-\frac{m_i}{\textup{Var}(Y_i)}\right)
\]
where
\[
\textup{Var}(Y_i)=\sum_{j=1}^N \frac{(Y_{i,j}-Y_{i,j-1})^2}{t_j-t_{j-1}}
\]
gives an estimate on the order of magnitude of the quadratic variation of the $i^{\textup{th}}$ component of the trajectory $x$. The prior for the topology was $p(S)=0.01^{|S|_0}$. Parallel tempering scheme was used in the simulations with 16 temperatures forming a geometric sequence $\beta_j^{-1} = 1.05^{j-1}$. The MCMC samples were initiated with a burn-in of 3000 samples. After that, every 10th sample was taken into the chain, which in the end consisted of 50000 samples in each case. In the parallel tempering scheme, swaps between states were attempted every 10th iteration. Every other time, swaps 1$\leftrightarrow$2, 3$\leftrightarrow$4,...,15$\leftrightarrow$16 were attempted, and every other time 2$\leftrightarrow$3, 4$\leftrightarrow$5,...,14$\leftrightarrow$15. In the heuristic tempering scheme implementation, the temperature $\beta^{-1}=1.5$ was used.

The method was compared to the EM-LASSO algorithm, where the E-step consists of computing the Kalman smoother (fixed lag 2) solution $\hat x^{(l)}$ (discretised) using the matrix $A^{(l)}$ from the previous iteration, and the M-step consists of solving the new matrix $A^{(l+1)}$ as the solution of the convex optimization problem
\[
\min_A \int_0^T \norm{\frac{d}{dt} \hat x^{(l)}-A\hat x^{(l)}}^2 dt + \lambda \norm{A}_1.
\]
In the case of two time series, the integrals corresponding to both trajectory estimates are combined to one cost function. Each case was tried with six different penalty values, $\lambda \in \{ 0.8, \ 1, \ 1.25, \ 1.5, \ 2, \ 3 \}$. Somewhat unfairly, the value producing the best results was chosen for comparison in each case.

\begin{table}[t]
\small
  \centering
\vspace{-2mm}
\caption{Classification scores in three different cases. Cases 2 and 3 were not tried with the parallel tempering scheme.}
    \vspace{2mm}
  \begin{tabular}{|l|ll|ll|ll|}
    \hline
    \hline
 & Parallel tempering \hspace{-20mm} & & Heuristic tempering \hspace{-20mm} & & EM-Lasso \hspace{-30mm} &  \\
  %& error & error & error \\
%\hline 
 &  {\footnotesize AUROC} &  {\footnotesize AUPREC} & {\footnotesize AUROC} &  {\footnotesize AUPREC} &  {\footnotesize AUROC} &  {\footnotesize AUPREC}  \\ 
 \hline
Case 1 & .9970 & .9414 & .9987 & .9766 & .9414 & .7494 \\
Case 2 & N/A & N/A & .9968 & .9588 & .9465 & .7788 \\
Case 3 & N/A & N/A & .8857 & .3984 & .8564 & .5626 \\
   \hline
    \hline
  \end{tabular}
  \label{tab:results}
\end{table}

%Altogether ten replicates were simulated using the same system with different initial conditions and noise realizations. The parallel tempering scheme was only applied on one of the replicates. For the dataset of the ten replicates, we instead used the heuristic tempering scheme described in Section~\ref{sec:temp}. The methods were compared using two standard classifier scores, the area under the receiver operating characteristic curve (AUROC) and the area under the precision recall curve (AUPREC). For the replicate for which parallel tempering was used, these values were ... and ..., respectively, while the heuristic tempering scheme gave ... and .... For the full dataset, the heuristic tempering scheme gave AUROC ...$\pm$... (mean $\pm$ standard deviation), and AUPREC. The scores for EM-Lasso were ... and ...

%It should be noted that a straightforward comparison of the two methods is not entirely meaningful. The posterior for $S$ should actually be considered as a multidimensional Bernoulli distribution, whose properties are not entirely reflected in the probabilities with which the components are non-zero. Indeed, the MCMC samples $S^{(l)}$ contain more information than just the probabilities revealed by the matrix $P$. This additional information can consist of --- for example --- joint probabilities $Pr(A_{i,j} \ne 0 \, \& \, A_{k,l} \ne 0)$. However, it should be noted that in an $n$-dimensional system there are already $n^4$ such joint probabilities of two possible links, so with larger problems it is not feasible to collect all such probabilities.
\begin{figure}[t]
\center
\mbox{\hspace{-1.5cm} \includegraphics[width=15cm]{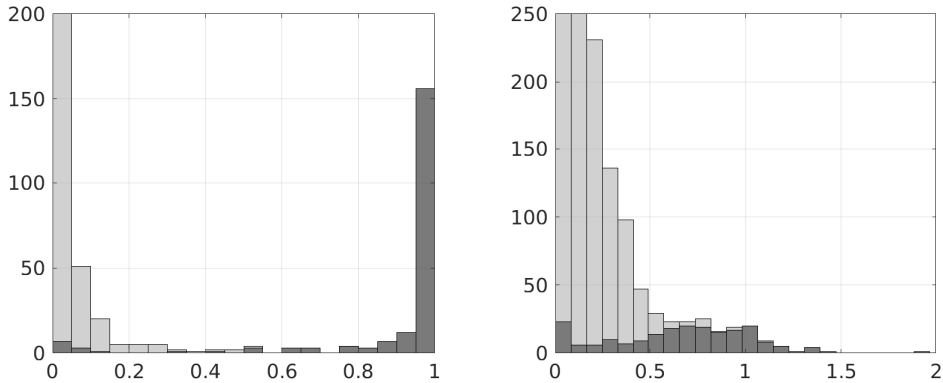}}
%\vspace{-1.5cm}
\caption{Results on Case 1 as stacked bar charts. Darker bars represent the entries that are non-zero in the true $A$.  Left: Results from the presented method with parallel tempering. X-axis shows probabilities of entries being non-zero. The first bar extends to 9715. Right: Results from the EM-Lasso. X-axis shows absolute magnitudes of the estimated matrix entries. First two bars extend to 8885 and 427.}
\label{fig:hist}
\end{figure}

\begin{figure}[t]
\center
\includegraphics[width=12.5cm]{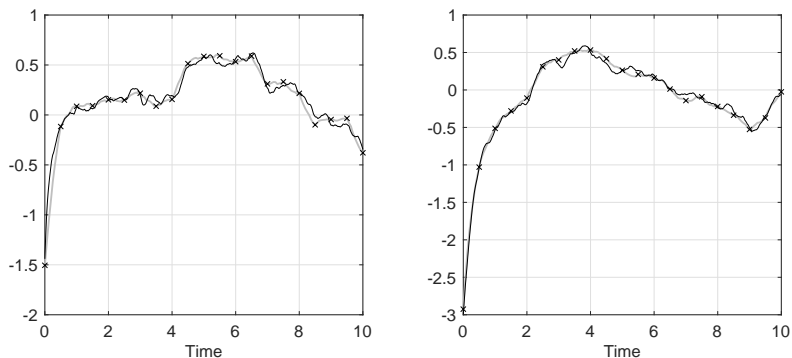}
%\vspace{-1.5cm}
\caption{One variable of the two time series. The noisy samples are marked with crosses, the true trajectory is shown with the black line and the trajectory estimate is shown with the thicker gray line.}
\label{fig:series}
\end{figure}

Three different cases were tried to compare the introduced method using both the parallel tempering scheme and the heuristic tempering scheme introduced in Section~\ref{sec:temp}, with the EM-Lasso:

\begin{itemize}
\item Case 1: Two time series with sampling time 0.5.
\item Case 2: Two time series with sampling time 1.
\item Case 3: One time series with sampling time 0.5.
\end{itemize}
The parallel tempering scheme was tried on the first case only. The methods were compared using two standard classifier scores, the area under the receiver operating characteristic curve (AUROC) and the area under the precision recall curve (AUPREC). These values are shown in Table~\ref{tab:results}. Some more illustration on Case~1 are shown in Figure~\ref{fig:hist} presenting stacked bar charts on the results of the parallel tempering and the EM-Lasso methods. It should be noted that the results of the presented method consist of probabilities for each entry in $A$ being non-zero. The EM-Lasso on the other hand, gives estimates of the magnitudes of the entries of the $A$ matrix. Nevertheless, it can be concluded that from the noisy data, the presented method is still able to do almost perfect job in this test problem with two time series. With threshold 0.5, the method finds 191 out of 204 true links with only one false positive. In case 3, the EM-Lasso achieved higher precision score, although by adjusting the topology prior to $p(S)=0.04^{|S|_0}$, the scores for the heuristic tempering scheme increased to 0.8890 and 0.5212. The optimal values for $\lambda$ in the three cases were 2, 1, and 0.8, respectively.
Figure~\ref{fig:series} shows one variable in the data to give an idea of its behavior. In addition, the true continuous trajectory and the conditional mean of the trajectory estimated using our method are shown in the plot.

%Three experiments were run with the following specifications
%\begin{itemize}
%\item[(i)] One time series with 
%\item[(ii)] Two time series with different initial conditions, otherwise as in (i).
%\item[(iii)] Two time series with both measurement and process noises multiplied by 2.
%\end{itemize}

\section{Discussion}

We presented a fully probabilistic method for variable selection in linear dynamical systems. The method performs well in our test problems. The method suffers somewhat from difficulties arising from sampling a combination of a discrete and a continuous variable. These difficulties are similar to multimodality problems in MCMC methods. They can be overcome by using a tempering scheme, such as the introduced parallel tempering scheme. However, it should be noted that a heuristic tempering approach seemed to work equally well, and with much smaller computational effort.

Compared to other types of sparse selection methods, a Bayesian MCMC approach provides more than a scored list of potential non-zero entries in the dynamics matrix. In fact, the posterior distribution for the zero-structure of the matrix $A$ should be considered as multivariate Bernoulli distribution \cite{Bernoulli}. The full multivariate Bernoulli distribution is characterized by $2^{n^2}$ parameters, which is clearly infeasible. However, some additional statistics besides $\Ex (S|Y)$ can be obtained from the collected MCMC samples.

Our further research includes incorporation of nonlinear dynamics by introducing dynamics functions modelled as Gaussian processes, whose covariance hyperparameters reveal the interconnection structure.

%\begin{itemize}
%\item Talk about the heuristic temperature scheme.
%\item Bernoulli distribution for $S$
%\item Nonlinear stuff using libraries as in  \cite{dyn_model_selection} \cite{SINDy}
%\end{itemize}

\bibliographystyle{plain}
\bibliography{bibli}

\end{document}